\documentclass[aps,prc,superscriptaddress,amsmath,showpacs,preprint,tightenlines,nofootinbib]{revtex4-2}

\usepackage[ngerman,english]{babel}
\usepackage{graphicx}
\usepackage{float}
\usepackage{mathtools}

\usepackage{physics}
\usepackage{nicefrac}
\usepackage{orcidlink}

\usepackage{ltpc}

\renewcommand{\i}{\mathrm{i}}
\renewcommand{\v}[1]{\ensuremath{\boldsymbol{#1}}}

\hypersetup{
    pdftitle={Triton neutron-neutron distribution in pionless EFT},
    pdfauthor={Tanja Kirchner, Matthias Göbel, Hans-Werner Hammer},
    pdfview=FitB
    pdfstartview=FitB
}

\usepackage{cleveref}

\begin{document}

    \title{\texorpdfstring{Neutron-neutron distribution of the triton from pionless EFT}{Neutron-neutron distribution in the triton from pionless EFT}}    
    
    \author{Tanja Kirchner\,\orcidlink{0009-0000-3735-058X}}
    \email{tanja.kirchner@tu-darmstadt.de}
    \affiliation{Technische Universität Darmstadt, Department of Physics, Institut f\"ur Kernphysik, 64289 Darmstadt, Germany}

    \author{Matthias Göbel\,\orcidlink{0000-0002-7232-0033}}
    \email{matthias.goebel@pi.infn.it}
    \affiliation{Istituto Nazionale di Fisica Nucleare, Sezione di Pisa, Largo Pontecorvo 3, 56127 Pisa, Italy}

    \author{Hans-Werner Hammer\,\orcidlink{0000-0002-2318-0644}}
    \email{Hans-Werner.Hammer@physik.tu-darmstadt.de}
    \affiliation{Technische Universität Darmstadt, Department of Physics, Institut f\"ur Kernphysik, 64289 Darmstadt, Germany}
    \affiliation{Helmholtz Forschungsakademie Hessen f\"ur FAIR (HFHF)
    and ExtreMe Matter Institute EMMI, GSI Helmholtzzentrum f\"{u}r Schwerionenforschung GmbH, 64291 Darmstadt, Germany}

    \date{May 27, 2025}

\begin{abstract}

We compute the neutron-neutron relative-energy distribution of the triton following the hard knockout of the proton in pionless effective field theory.
This distribution can be used to study universality as well as to obtain information on the neutron-neutron interaction.
Especially, one can infer the scattering length from fitting theory predictions for the shape of the distribution to experimental data.
To obtain the distribution for the triton, we first solve the ground-state three-body problem using momentum-space Faddeev equations.
Next, we include the neutron-neutron final-state interaction by applying the corresponding M{\o}ller operator to the ground state.
We present leading-order (LO) and next-to-leading order (NLO) pionless effective field theory results with quantified
uncertainties.
At NLO, we include the effective ranges semi-perturbatively.
We conclude, that 
pionless EFT works reliably as expected and that the neutron-neutron distribution of the triton shows a significant sensitivity to the scattering length.

\end{abstract}

\maketitle
\newpage

\section{Introduction}

The neutron-neutron relative-energy distribution of weakly bound nuclear systems following the sudden knockout of the remaining core of the nucleus offers important insights into the structure of the nucleus and the neutron-neutron interaction.
To date, it has been investigated for two-neutron halo nuclei, i.e. nuclei consisting of a tightly bound core and two weakly bound neutrons. G\"obel and collaborators have studied the distribution to analyze the degree of universality of two-neutron halo nuclei \cite{Gobel:2023rpj}. They showed that the distributions obtained in halo effective field theory (see, e.g., Refs.~\cite{Hammer:2017tjm,Hammer:2019poc,Hammer2023} for reviews) indeed fall on a universal curve. Moreover, since the distribution is strongly influenced by the neutron-neutron final-state interaction, it has been proposed as an observable suitable to determine the neutron-neutron scattering length \cite{nn_scat_len_ribf_prop2018}. The neutron-neutron scattering length characterizes the fundamental neutron-neutron interaction at threshold. Therefore, it is an important parameter for many calculations, either explicitly or implicitly, e.g. in the construction of chiral effective field theory interactions \cite{Bedaque:2002mn,Epelbaum:2008ga,Machleidt:2011zz,Hammer:2019poc}. Additionally, it can also be used to characterize the amount of charge-symmetry breaking in nuclear forces (see, e.g., Ref.~\cite{Miller:2006tv}). The fundamental importance of the neutron-neutron scattering length is not matched by the precision of its experimental value, as there is some tension between different experimental determinations (see, e.g., Refs. \cite{Gobel:2021pvw,Gardestig:2009ya} for discussions).

In order to resolve this unsatisfactory situation, 
a new measurement with very different systematics was proposed by Aumann et al.~\cite{nn_scat_len_ribf_prop2018} and will be carried out at RIKEN/RIBF. The main idea is to determine the scattering length by fitting theory predictions for the distribution parameterized by the neutron-neutron scattering length to the experimental data for the reactions $^6$He($p,p\alpha$)$nn$ and $t(p,2p)nn$ in inverse kinematics. The knockout is done with a high momentum transfer to cause a large kinematical separation of the two neutrons and the knocked-out core. As a result, final-state interactions (FSIs) between the neutrons and the core are strongly suppressed, allowing for a clean imprint of the \(nn\) interaction on the measured distribution. Since the two neutrons have a low relative energy and a large center-of-mass energy, each one has almost the same absolute energy. Thus the influence of the energy-dependent neutron detection efficiency is negligible. For more details see Ref.~\cite{nn_scat_len_ribf_prop2018}. 

The  \(nn\) distribution for the case of \(^6\)He was calculated in Ref.~\cite{Gobel:2021pvw} using halo effective field theory. The methodology for including the \(nn\) final-state interaction and for obtaining the distribution was discussed in detail.
A variation of the neutron-neutron scattering length $a_{nn}$ by 2 fm around the central value of -18.7 fm causes a change in the peak-to-tail ratio of the \(E_{nn} \leq 1\) MeV part of the distribution by approximately 10 \%. Thus, a measurement of the distribution provides a viable method to extract $a_{nn}$.
Recently, the two-neutron distribution in $^6$He($p,p\alpha$)$nn$ was measured as a benchmark for verifying the analysis and calibration procedures in the four-neutron experiment of
Ref.~\cite{Duer:2022ehf} and shown to agree with the theoretical predictions. However, the precision in that experiment was not sufficient to extract the scattering length.

The purpose of this work is to calculate the neutron-neutron relative-energy distribution for the hard 
knock-out reaction $t(p,2p)nn$ in inverse kinematics.
The general procedure is similar as in Ref.~\cite{Gobel:2021pvw}. First, the ground state is described by solving the three-body momentum-space Faddeev equations.
Afterwards, the action of the M{\o}ller operator describing \(nn\) FSI on the ground-state wave function is evaluated and finally the distribution is obtained from the wave function.
The calculation is carried out in the framework of pionless effective field theory (pionless EFT) \cite{vanKolck:1997ut,Kaplan:1998tg,Kaplan:1998we,vanKolck:1998bw,Chen:1999tn} as protons and neutrons are the appropriate degrees of freedom for the triton.
Pionless EFT provides a comprehensive description of the bound state and scattering
properties of the triton~\cite{Bedaque:1999ve,Bedaque:2002yg,Vanasse:2013sda,Vanasse:2015fph,Vanasse:2017kgh}.
The interactions between the nucleons are specified in terms of dimer propagators or equivalently t-matrices with denominators constrained by the effective-range expansion. The power counting scheme of pionless EFT specifies which terms of the effective-range expansion are contained in the t-matrices to a given order in the EFT expansion. Further details on pionless EFT can be found in the reviews~\cite{Bedaque:2002mn,Epelbaum:2008ga,Hammer:2019poc}.
In Refs. \cite{Bedaque:1999ve,Bedaque:2002yg,Vanasse:2013sda,Vanasse:2015fph,Vanasse:2017kgh}, the calculations were implemented using a field-theoretical formalism for the scattering amplitudes. 
In this work, we employ the same power counting but use the equivalent formulation of the momentum-space Faddeev equations for the wave function (see, e.g., Ref.~\cite{Gobel:2024ovk} for details).

This paper is structured as follows:
First, we lay out the methodology. We start by discussing the basics of the ground-state calculations. We continue with the specific details of the leading-order and the next-to-leading-order calculation. The exposition of the methodology concludes with the inclusion of final-state interactions.
Then, we show our results. We begin with ground-state distributions, going from leading to next-to-leading order. Finally, we discuss the results including final-state interactions. In particular, we analyze their sensitivity on the neutron-neutron scattering length.

\section{Formalism and methodology}
In this section, we present our formalism and methodology. A more detailed discussion can be found in Ref.~\cite{Gobel:2024ovk}.
First, the focus is on the ground-state calculation in the Faddeev formalism and then we move on to the calculation of final-state interactions.
The computations are carried out in momentum space. We use  the Jacobi basis for the momenta.
This is a basis of relative momentum vectors \(\v{p}_i\) and \(\v{q}_i\), which are defined with respect to a spectator \(i\) in a system of three particles \(ijk\) (\(i\neq j,\, j\neq k,\, k \neq i\)).
The momentum \(\v{p}_i\) is then the relative momentum between the particles \(j\) and \(k\).
The vector \(\v{q}_i\) specifies the momentum between particle \(i\) and the center of mass of \(j\) and \(k\).
The definition of the Jacobi momenta in terms of the single-particle momenta \(\v{k}_i\), \(\v{k}_j\), and \(\v{k}_k\) reads
\begin{align}
    \v{p}_i &\coloneqq \mu_{jk} \K{\frac{\v{k}_j}{m_j} - \frac{\v{k}_k}{m_k}} \,, \\
    \v{q}_i &\coloneqq \mu_{i(jk)} \K{ \frac{\v{k}_i}{m_i} - \frac{\v{k}_j + \v{k}_k}{M_{jk}} }\,,
\end{align}
whereby the single-particle masses are given by the \(m_i\).
The relevant mass factors are defined by \(\mu_{ij} = m_i m_j / (m_i + m_j)\), \(M_{ij} = m_i + m_j\), and \(\mu_{i(jk)} = m_i M_{jk} / (m_i + M_{jk} )\).
The momenta usually appear within bra or ket states. To simplify the notation, we omit the index indicating the spectator at the momenta and use a single index to the left of the bra or the right of the ket state instead. This reads for example as \(\iket{p,q}{n}\) for the neutron as spectator.
Inter alia, this notation has the advantage of reducing the number of indices written.

We work in a partial-wave basis, because the interactions are dominated by their components in a few partial waves.
The orbital-angular-momentum quantum number in the \(jk\) subsystem is denoted by \(l\), the one for the motion of particle \(i\) relative to \(jk\) is denoted by \(\lambda\). The overall spin of particles \(j\) and \(k\) is \(s\), the total angular momentum in that subsystem is \(j\).
Particle \(i\) has the spin quantum number \(\sigma\) and from the coupling of \(\lambda\) and \(\sigma\) emerges \(I\).
The overall angular momentum/spin of the three-particle system is denoted by \(J\) with projection \(M\).
It is convenient to bundle the quantum numbers in a multiindex \(\Omega\) .
In \(jI\) coupling \(\Omega\) reads
\begin{equation}\label{eq:Omega}
    \Omega = \K{l,s}j \K{\lambda, \sigma} I; J, M \,.
\end{equation}
Finally note that the quantum numbers defined here require the notion of the spectator particle.

\subsection{Faddeev equations}

In order to obtain the bound state of the triton in pionless EFT we use the momentum-space Faddeev formalism.
In this formalism the interactions are represented in terms of t-matrices, which is advantageous for our EFT treatment.
The power counting tells us which terms in the on-shell t-matrix's denominator are contained.
At leading order, the $S$-wave neutron-neutron and neutron-proton scattering lengths and a three-body parameter, usually fixed to the triton binding energy, contribute.
At next-to-leading order the effective ranges enter as perturbative corrections.
The Faddeev equations are a system of coupled equations determining the Faddeev amplitudes \(\ket{F_i}\), which are related to the
overall state \(\ket{\Psi}\) by 
\begin{equation}\label{eq:def_F_i}
  G_0 t_i \ket{F_i} = \ket{\psi_i} = G_0 V_i \ket{\Psi} 
\end{equation}
with the Faddeev state component \(\ket{\psi_i}\).
The interaction between particles \(j\) and \(k\), i.e., in the \(jk\) subsystem, is specified by the spectator index \(i\) and the t-matrix \(t_i\).
The Faddeev equations read
\begin{equation}\label{eq:Fd}
  \ket{F_i} = \sum_{j \neq i} G_0 t_j \ket{F_j} \,. 
\end{equation}
In the simplest case there is one interaction channel with a specific set of quantum numbers \(\Omega_i\) and one t-matrix \(t_i\) per spectator \(i\).
However, for example in the case of the triton for the \(np\) system there are two interaction channels at leading order:
the singlet as well as the triplet channel.
In that case, one needs to extend the indexing scheme to account for that.
Basically, there are two ways to realize this. One is to add more indices and the other is to change the meaning of the index.
We choose the latter option and accordingly Eq.~(\ref{eq:Fd}) requires no modification.
The index \(i\) is now just a numeric index and can take as many values as different interactions are present. In our case, there are three values: 0 for the \(nn\) system, 1 for the \(np\) singlet system,
and 2 for the \(np\) triplet system, which supports the deuteron ($d$) as a bound state.
In this case the spectator information is no longer contained in the index \(i\) itself.
However, one can introduce the function \(\mathcal{S}{(i)}\) for storing the spectator information.
The Faddeev formalism as presented in the equations above is formulated for distinguishable particles.
In principle this implies five interaction channels and correspondingly five Faddeev amplitudes.
These are one \(nn\) channel, two times the \(np\) spin singlet channel (differing in the involved neutron), and two times the \(np\) spin triplet.
The antisymmetrization due to the indistinguishability of the two involved neutrons relates the two \(np\) spin singlet channels
with each other based on the \(nn\) permutation operator \(\pmo\). The same holds for the two \(np\) spin triplet channels.
Antisymmetrization can now be fulfilled by using these five interactions obeying the discussed relations.
Alternatively, one can explicitly use the symmetry to introduce revised Faddeev equations for the three independent amplitudes.
This approach is employed here.
The interactions \(i\) involving one neutron and their relatives are related by \(t'_i = \K{-\pmo} t_i \K{-\pmo}\).
Making use of this, the Faddeev equations given in \Cref{eq:Fd} change into
\begin{equation}\label{eq:Fd_as}
  \ket{F_i} = \sum_{j=0}^2 G_0 A_{ij} t_j \ket{F_j} \,,\quad i=0,1,2\,,
\end{equation}
whereby \(A_{ij}\) introduces the extra-terms due to antisymmetrization for the \(j\) with \(\mathcal{S}{(j)}=n\), i.e., the terms with interaction
in the \(np\) system.

In order to solve the Faddeev equations one needs a representation of the Faddeev amplitudes.
For separable interactions it is advantageous to use the representation
\begin{equation}\label{eq:def_F_i_q}
  F_i{(q)} \coloneqq \rint{\p} g_i{(p)} \ibraket{\mathcal{S}{(i)}}{p,q; \Omega_i}{F_i}{} \,,\quad i=0,1,2\,,
\end{equation}
where the \(g_i\) are the form factors of the separable interactions.
Based on this representation Eq.~(\ref{eq:Fd_as}) then reads
\begin{equation}\label{eq:Fd_as_rep}
  F_i{(q)} = 4\pi \sum_{j=0}^2 \rint{\qp} X_{ij}{\K{q, \qp; E_3}} \tau_j{\K{\qp; E_3}} F_j{\K{\qp}} \,,\quad i=0,1,2\,,
\end{equation}
with the exchange kernel
\begin{equation}\label{eq:def_X_ij}
  X_{ij}{\K{q, \qp; E_3}} \coloneqq \rint{\p} \rint{\pp} g_i{\K{p}} \imel{\mathcal{S}{(i)}}{p,q;\Omega_i}{G_0{\K{E_3}} A_{ij}}{\pp,\qp;\Omega_j}{{\mathcal{S}{(j)}}} g_j{\K{\pp}} \,.
\end{equation}
The \(\tau_i{\K{q; E_3}}\) is the three-body embedding of the reduced t-matrix \(\tau_{i}{\K{E_2}}\):
\(\tau_i{\K{q; E_3}} \coloneqq \tau_{i}{\K{E_3 - q^2 / \K{2\mu_{i(jk)}}}}\).
Explicit expressions for the exchange kernel for interactions in arbitrary partial waves can be found in chapter 9 of Ref.~\cite{Gobel:2024ovk}.
Additionally, a three-body force is included in the Faddeev equations to renormalize the three-body energy.
The triton binding energy serves as renormalization condition.
The force is introduced in the Faddeev equations (Eq. \ref{eq:Fd_as_rep}) by replacing \(X_{00}{\K{q,q'}}\) by \(X_{00}{\K{q,q'}} + h{(\Lambda)}\) and \(X_{11}{\K{q,q'}}\) by \(X_{11}{\K{q,q'}} + h{(\Lambda)}\). Hereby, \(h{(\Lambda)}\) denotes the running three-body coupling.
It is not dimensionless, but can be rescaled to a dimensionless coupling.

Given that the Faddeev formalism can not only be derived from the Schrödinger equation but also from the Feynman diagrams describing the three-body dynamics,
the concrete equations originating from the presented formalism have to be the same as the one presented in the field-theoretical treatment of the triton in pionless EFT by Bedaque et al. in Ref.~\cite{Bedaque:2002yg}.
We explicitly derived the concrete form of the equations resulting from the formalism presented above and verified the consistency with the equations of Ref.~\cite{Bedaque:2002yg} for the case of a cutoff with a vanishing three-body force\footnote{
    In the isospin-symmetric case the relations between the amplitudes of Bedaque \textit{et al.} (\(F^{(0)}\) and \(F^{(1)}\)) and ours are \(F^{(1)} = \sqrt{3} F_2\) and \(F^{(0)} = F_1 - F_0\).
}. Note that due to the limit cycle behavior of the (dimensionless) three-body force, such a cutoff can always be found.
For the numerical calculations, we use a computer implementation of the formalism and the computer code automatically constructs the explicit equation system on the basis of the interactions and expressions for \(X_{ij}\) for arbitrary partial waves.

In principle one can assemble expressions for the observables directly from the Faddeev amplitudes.
However, to reduce the effort and to modularize the subsequent calculations, one can use the full wave function of the bound state.
The overall wave function as seen from spectator \(\mathcal{S}\) in partial wave \(\Omega\) is given by
\begin{equation}\label{eq:gs_wf}
  \Psi_{\mathcal{S};\Omega}{\K{p,q}} \coloneqq \ibraket{\mathcal{S}}{p,q;\Omega}{\Psi}{} \,.
\end{equation}
In order to obtain the wave function from the Faddeev amplitudes, one uses \(\ket{\Psi} = \sum_i \ket{\psi_i}\) in combination with Eq.~(\ref{eq:def_F_i}) and its representations (its projections on basis states).
An explicit expression for the overall wave function is given by Eq. (9.20) in Ref. \cite{Gobel:2024ovk}.
For the evaluation it is helpful to make use of Eq. (9.101) of the same reference.

\subsection{Ground-state neutron-neutron distributions}

From the ground-state wave function as given in Eq.~\eqref{eq:gs_wf} one can obtain the ground-state \(nn\) relative-momentum distribution \(\rho_{nn}{(p)}\) in the following way:
\begin{equation}
    \rho_{nn}{(p)} = \sum_{\Omega} \int \dd{q} p^2 q^2 \left| \Psi_{c;\Omega}{\K{p,q}} \right|^2 \,,
\end{equation}
whereby contributions from all the different partial waves \(\Omega\) are added.
In practice, it turns out that the component resulting from the partial-wave state
\begin{equation}
    \iket{\Omega^{(0)}}{c} = \iket{(0, 0)0, (0, \frac{1}{2})\frac{1}{2};\frac{1}{2},M}{c}
\end{equation}
is by far the most important.
For simplicity, we define \(\Psi_{c}{\K{p,q}} \coloneqq \Psi_{c;\Omega^{(0)}}{\K{p,q}}\).

Instead of using the \(nn\) relative-momentum distribution one can also use the \(nn\) relative-energy distribution.
It is obtained from the momentum distribution via
\begin{equation}
    \label{eq:rho_E_rho_p}
    \rho_{nn}{\K{E}} = \sqrt{\frac{\mu_{nn}}{2E_{nn}}} \rho_{nn}{\K{\sqrt{2\mu_{nn} E_{nn}}}} \,.
\end{equation}
This relation follows from the normalization relations $\int \dd{p} \rho_{nn}{\K{p}} = 1$ and
$\int \dd{E_{nn}} \rho{\K{E_{nn}}} = 1$ via integration by substitution.

\subsection{Two-body t-matrices}
\label{sec:semipert}

At LO, the two-body t-matrices $t_i$ with  $i=0,1$, representing the $nn$ and spin-singlet $np$ channels, are determined by the 
corresponding scattering lengths. In the 
spin-triplet $np$ or deuteron ($d$) channel ($i=2$), 
the deuteron binding momentum enters instead.
At NLO, the corresponding effective ranges contribute as well.
In the $nn$ and $np$ singlet channel, the scattering lengths are negative. Here, we use the standard effective range expansion around the scattering threshold. In the deuteron ($d$) channel, a modified effective range expansion of the t-matrix around the bound state pole is more convenient.

After including the effective range as a second order term in the denominator, the reduced t-matrix, 
\begin{equation}
    \tau_i(p) = \frac{-1}{4\pi^2 \mu_{jk}} \frac{1}{-1/a_i + {r_i}p^2/2 - \i p}\,, \quad i=0,1
\end{equation}
with $p=\sqrt{2\mu_{jk}E}$,
develops a spurious pole at $p\sim 1/r_i$. 
This unphysical pole is at the breakdown scale of pionless EFT and does not correspond 
to a physical bound state. It leads to problems with unitarity in three-body calculations. We thus follow the treatment of Ref.~\cite{Bedaque:2002yg} and
expand the reduced t-matrix in the effective range term $r_i p^2/2$ up to first order:
\begin{equation}
\label{eq:semipert}
\frac{1}{-1/a_i + {r_i}p^2/2 - \i p}
=\frac{1}{-1/a_i - \i p}
\left(1-\frac{r_i p^2/2}{-1/a_i-\i p}+\ldots  \right)
\,, \quad i=0,1
\end{equation}
With this expansion, the spurious pole is no longer present.

In the deuteron channel ($i=2$) we apply
the reduced t-matrix,
\begin{equation}
    \tau_{2}(p) = \frac{-1}{4\pi^2 \mu_{nn}} \frac{1}{-\gamma_{d} +\rho_{d}({\gamma_{d}\,}^2 + p^2)/2 -\i p}
\,.    \label{eq:polexpansion}
\end{equation}
It uses the deuteron binding momentum $\gamma_{d}$ and the effective range $\rho_{d}$.\footnote{To this order the effective ranges $r_i$ and $\rho_i$ from the expansions around threshold and around the bound state pole can be taken to be the same.} This reduced t-matrix also exhibits the spurious deep pole and is expanded in the same way as Eq.~\eqref{eq:semipert}.

Using these ``partially resummed'' t-matrices includes all terms required by the power counting at NLO, but generates some higher-order terms in the three-body system. These higher-order contributions have to be
small for a consistent calculation. We note that this requirement puts some restrictions on the values of the cutoff $\Lambda$ which can not be chosen too large. In our calculations below, we choose $\Lambda = 700$ MeV, where the calculations are well converged.
A fully perturbative treatment of the effective range corrections that only includes the next-to-leading order contributions is also possible \cite{Hammer:2001gh,Ji:2011qg,Ji:2012nj,Vanasse:2013sda}
but is not considered here.

So far, we have mainly discussed the momentum dependence of the interactions. We complement this with reviewing their partial-wave structure, following the nomenclature of Eq.~(\ref{eq:Omega}).
The \(nn\) as well as the spin-singlet \(np\) interaction act in the partial wave \((0,0)0\), where we use the quantum number order \((l,s)j\).
If we collect these quantum numbers in the multiindex \(\omega\), we have \(\omega_0 = \omega_1 = (0,0)0\).
The interaction channels are then given by the projection operators \(\iketbra{\omega_0}{c}{}\) and \(\iketbra{\omega_1}{n}{}\), respectively.
The spin-triplet \(np\) interaction takes place in \(\omega_2 = (0,1)1\).
The projection operator is \(\iketbra{\omega_2}{n}{}\).
For the three-body calculations it is quite advantageous to check if these subsystem quantum numbers determine also the other quantum numbers of the three-body system for the nucleus considered.
Since the triton has the quantum numbers \(J^\pi = 1/2^+\), we must have \(J=1/2\) in Eq.~\eqref{eq:Omega}.
Moreover, 
the third particle is always a nucleon of spin \(1/2\) and we have \(\sigma = 1/2\).
For the \(nn\) channel as well as for the \(np\) spin-singlet channel, we have now \(j=0\) and \(J=1/2\).
Angular momentum coupling implies that \(I\) must be \(1/2\) and, consequently, \(\lambda\) must be either \(0\) or \(1\).
The overall positive parity together with \(l=0\) excludes the latter possibility yielding
\begin{equation}
  \Omega_0 = \Omega_1 = (0,0)0 (0,\frac{1}{2})\frac{1}{2}; \frac{1}{2}, M \,.
\end{equation}
While the multiindices themselves are the same, the projection operators based on them are of course different. This is because the channels themselves are not the same, but have different particle content.
The operators are \(\iketbra{\Omega_0}{c}{}\) and \(\iketbra{\Omega_1}{n}{}\).
For the \(np\) spin-triplet from \(j=1\) and \(J=1/2\) we conclude that \(I\) must be either \(1/2\) or \(3/2\).
Taking also the parity constraint together with \(l=0\) into account, we have \(\lambda = 0\) for \(I=1/2\) and \(\lambda=2\) for \(I=3/2\).
For simplicity, we neglect the second configuration which is suppressed at low energies. However, we stress that the employed framework can handle this component without changing the formalism. In the current formalism it would simply give rise to a fourth Faddeev component (\(i=3\)) with an \(\Omega_3\) having \(\lambda=2\) and \(I=3/2\).
Accordingly, we have
\begin{equation}
  \Omega_2 = (0,1)1 (0, \frac{1}{2})\frac{1}{2}; \frac{1}{2}, M 
\end{equation}
with the projection operator reading \(\iketbra{\Omega_2}{n}{}\).

\subsection{Final-state interactions (FSI)}
\label{sec:FSI}

After the knockout of the proton from the triton, the so-called final-state interactions (FSIs) of the  fragments occur.
Due to the kinematical condition of a hard knockout yielding a strong separation of the neutrons from the proton in momentum space, final-state interactions are dominated by the \(nn\) interaction.
Corrections from other final-state interactions are suppressed by \(p/k\), where \(k\) is the large momentum transfer to the proton in the knockout and $p$ is the relative momentum of the neutrons.
The strong separation in momentum space corresponds to a large \(k\).

The \(nn\) FSI is included on top of the ground-state Faddeev calculations following Ref.~\cite{Gobel:2021pvw}. We consider two treatments: (i) an exact calculation of the 
\(nn\) FSI which determines the FSI on the basis of the ground-state wave function as well as the
neutron-neutron t-matrix and
(ii) and approximate calculation based on 
an FSI enhancement factor which scales the final probability distribution for the ground state.
We use the exact method (i) for our main results. 
In Appendix~\ref{ap:FSI_tm_ef}, we compare both methods
and show that the approximate enhancement 
factor method agrees well with the exact 
method at NLO.

\subsubsection{t-matrix method}

On the basis of the ground-state wave function $\Psi_c(p, q)={}_c\braket{p, q;\Omega_c^{(0)}}{\Psi}$ obtained in the momentum-space Faddeev formalism discussed above, one can calculate the wave function after final-state interactions between the two neutrons from the $nn$ t-matrix as follows:
\begin{align}
\label{eq:exactFSI}
    \Psi_c^{(\mathrm{FSI})}(p, q) &= \Psi_c(p, q) + \frac{2}{\pi}g_0(p)\frac{1}{a_{nn}^{-1}-r_{nn}\,p^2/2+\i p} \\ \nonumber
    &\quad \times \qty[\int_0^\Lambda \dd p'\,\frac{p'^2\Psi_c(p', q)-p^2\Psi_c(p, q)}{p^2-p'^2}-\qty(\frac{\i\pi}{2}-\frac{1}{2}\ln\qty(\frac{\Lambda+p}{\Lambda-p}))g_0(p)\,p\,\Psi_c(p, q)]\ \,.
\end{align}
A detailed derivation of this formula can be found in Ref.~\cite{Gobel:2021pvw}.
The probability density \(\rho_{nn}^{(\mathrm{FSI)}}\) is obtained as
    \begin{equation}
    \label{eq:exactFSI-rho}
      \rho_{nn}^{(\mathrm{FSI})} (p)= \int \dd q\, p^2 q^2 \left|\Psi_c^{\K{\mathrm{FSI}}}{\K{p,q}}\right|^2 \,,
    \end{equation}
    while the relative-energy distribution can be calculated from this distribution by using Eq.~\eqref{eq:rho_E_rho_p}.

\subsubsection{Enhancement factors}
As explicitly demonstrated in Ref.~\cite{Gobel:2021pvw}
for the reaction $^6$He($p,p\alpha$)$nn$, the $nn$ FSI 
can be taken into account approximately by applying the enhancement factor
\begin{equation}
    G(p)=\frac{((p^2+\alpha^2)r_\mathit{nn}/2)^2}{(-1/a_{nn}+r_{\mathit{nn}}\,p^2/2)^2+p^2}\,,
    \label{eq:enhancementfactor}
\end{equation}
with $\alpha=(1+\sqrt{1-2r_{nn}/a_{nn}})/r_{nn}$ to the ground-state momentum distribution $\rho_{nn}$.
The neutron-neutron momentum distribution after FSI is then obtained via
\begin{equation}
\label{eq:enhancementfactor-rho}
    \rho_{nn}^{(\mathrm{FSI})}(p)\propto G(p) \rho_{nn}(p)\,.
\end{equation}
A detailed discussion of FSI and its treatment using enhancement factors can be found in Refs.~\cite{GoldbergerWatson64,Slobodrian:1971an}.
As Eq.~\eqref{eq:enhancementfactor} is not defined for vanishing effective ranges, the effective range $r_{nn}$ is always set to the physical value $r_{nn}=2.84\,$fm in the enhancement factor FSI calculations. However, the sensitivity to the exact value of $r_{nn}$ is small \cite{Gobel:2021pvw}.
Again, the conversion from a momentum to an energy distribution is given by Eq.~\eqref{eq:rho_E_rho_p}.

\section{Results}

We now discuss our results for the relative-energy distribution of the neutron in the hard knock-out reaction $t(p,2p)nn$.
The discussion is structured as follows.
We start with the leading-order results for the ground-state distribution, i.e., the distribution without FSIs.
Next we consider the ground-state distribution at next-to-leading order.
Finally, we present the NLO distribution with the \(nn\) final-state interaction.
Important aspects of the discussions are the sensitivity to variations of the \(nn\) scattering length as well as estimates of the EFT uncertainties.
Because the shape of the distribution will be used in the 
analysis of the planned experiment \cite{nn_scat_len_ribf_prop2018}, we always normalize the distribution to be one at a certain energy (chosen as $E_{nn}=0.8\,$MeV). 
For better readability, we refer to the different channels 
$i=0,1,2$ of the two-body t-matrices $t_i$
with their particle content as $nn$, $np$, and $d$.

\subsection{LO ground-state distribution}
We first specify the two-body interaction parameters used 
in our calculation. These parameters are listed in Tab.~\ref{tab:parameters_scatteringlength}.
\begin{table}[H]
    \centering
    \begin{tabular}{c|c|c}
         $a_{nn}$ [fm] & $a_{np}$ [fm]& $\gamma_{d}$ [MeV]\\
         \hline
        \{$-16.7, -18.7, -20.7, -23.714$\} & \,$-23.736$\, & \,$45.7021$\, \\
    \end{tabular}
    \caption{Two-body scattering lengths and deuteron pole momentum. The scattering lengths are given in fm and the deuteron pole is given in MeV. The value for the $np$ singlet-channel is taken from Ref. \cite{Reinert:2022jpu}.
    The scattering lengths for the $nn$ channel are varied to analyze the effect of the scattering length on the final distribution.}
    \label{tab:parameters_scatteringlength}
\end{table}
For the $nn$ scattering length, we compare four different values in order to quantify the sensitivity of the neutron distribution to $a_{nn}$.
The $np$ scattering length in the spin-singlet channel 
is taken from  the chiral EFT based analysis of Ref. \cite{Reinert:2022jpu}.
For the deuteron channel an effective range expansion around the pole is used, see Eq.~\eqref{eq:polexpansion}. The equation is given up to next-to-leading order in terms of the bound-state pole momentum with $B_2$ the deuteron binding energy and $\gamma_{d}=\sqrt{2\mu_{np}B_2}$ 
and the effective range $\rho_{d}$. At LO, $\rho_{d}\equiv 0$.
The three-body force in the pionless EFT calculation is 
chosen to reproduce the triton binding energy, $B_3=8.48\,$MeV.

\begin{figure}[htb]
    \centering
    \includegraphics[width=0.7\textwidth]{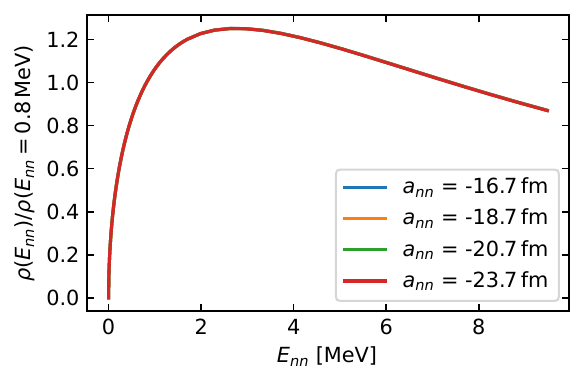}
    \caption{Ground-state neutron relative-energy distribution at LO for different values of $a_{nn}$ obtained with a cutoff of $\Lambda=700\,$MeV.
    All curves are on top of each other.}
    \label{fig:LOplot}
\end{figure}
The resulting ground-state distributions for \(E_{nn}\) of up to approximately 10 MeV are shown for different \(a_{nn}\) in Fig. \ref{fig:LOplot}. The distributions experience a steep rise at zero energies up to a peak at $E_{nn}=2.62\,$MeV (the same value for all four curves). For higher energies there is a slowly decaying tail.
Comparing the curves for the different values of the scattering length $a_{nn}$, we find that the position of the maximum does not change at all and only the value of the peak changes minimally (less than $0.03\%$ for each curve compared to the curve with $a_{nn}=-18.7\,$fm). We conclude that the dependence of the ground-state distribution on the scattering length is minimal. This observation is similar to the findings in the $^6$He case in Ref.~\cite{Gobel:2021pvw}. This insensitivity of the ground-state distribution to the exact value of $a_{nn}$ can be understood from the closeness of $a_{nn}$ to the unitary limit.
(See Ref.~\cite{Gobel:2023rpj} for more detailed discussion of this issue.)

\subsection{NLO ground-state distribution}

The next step is to go to NLO in the EFT expansion.
Our NLO results are obtained by treating the t-matrices in the Faddeev calculations semi-perturbatively as explained in \cref{sec:semipert}.
The values used for the effective ranges in the various channels are listed in Tab.~\ref{tab:parameters_effectivelength}.
The three-body force is 
adjusted to reproduce the triton binding energy, $B_3=8.48\,$MeV.
\begin{table}[H]
    \centering
    \begin{tabular}{c|c|c}
         $r_{nn}$ [fm] & $r_{np}$ [fm] & $\rho_{d}$ [fm] \\
         \hline
        \,$2.84$\, & \,$2.704$\, & \,$1.752$\, \\
    \end{tabular}
    \caption{Effective ranges for the two-body channels. All values are given in fm and taken from Ref. \cite{Reinert:2022jpu}.}
    \label{tab:parameters_effectivelength}
\end{table}

\begin{figure}[htb]
    \centering
    \includegraphics[width=0.7\textwidth]{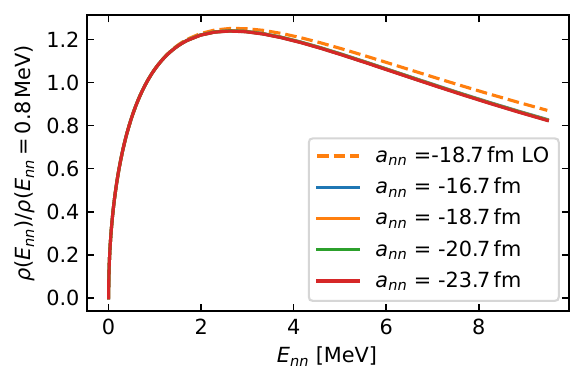}
    \caption{NLO results for the ground-state neutron relative-energy distribution obtained for a cutoff of $\Lambda=700\,$MeV.    The effective range parameters are given in Tab.~\ref{tab:parameters_scatteringlength} and Tab.~\ref{tab:parameters_effectivelength}
    while the three-body force is tuned to reproduce $B_3=8.48\,$MeV. For comparison, the LO ground-state distribution is shown in dashed.}
    \label{fig:NLOplot}
\end{figure}
The ground-state distribution up to NLO is shown in Fig.~\ref{fig:NLOplot}. We display the distribution up to energies of approximately $10\,$MeV. It is apparent that the investigated variations of the scattering length do not have a significant influence on the ground state distribution. Again, the position of the peak ($E_{nn}=2.62\,$MeV) does not change when $a_{nn}$ is varied and the peak height changes less than $0.1\%$ compared to the curve with $a_{nn}=-18.7$ fm. This is expected as the LO curve already had a negligible sensitivity to the scattering length and the relative effective range correction $r_{nn}/a_{nn}\approx -0.12\ldots -0.17$ varies only slightly for the considered values of $a_{nn}$.
Compared to the LO curve the peak position is unchanged. The overall shape of the NLO curve is very similar to the LO curve (shown as dashed).
The relative deviation of the NLO result from the LO one is less than 1\,\% at relative energies of 1 MeV and  less than 10\,\% at 10 MeV, and thereby smaller than the \textit{a priori} expectations from the power counting. 
At low energies, one expects the errors to be dominated
by the $r_{nn}/a_{nn}$ corrections while finite momentum corrections of order $\sqrt{2\mu_{nn}E_{nn}}/m_\pi$
take over at higher energies. Here the pion mass $m_\pi \sim 1/r_{nn}$ is taken as an estimate of the breakdown scale of the EFT.

\subsection{Relative-energy distribution with FSI}
\label{ssec:results_wfsi}

We now include the final-state interactions using the 
t-matrix method described in Sec.~\ref{sec:FSI}. 
The effective range parameters for the calculation of the triton ground-state wave function are given in Tab.~\ref{tab:parameters_scatteringlength} and Tab.~\ref{tab:parameters_effectivelength}. The three-body force is tuned to reproduce the triton binding energy $B_3=8.48\,$MeV. 

\begin{figure}[htb]
    \centering
    \includegraphics[width=0.7\textwidth]{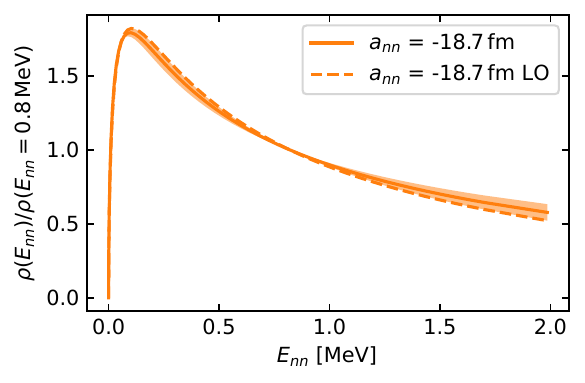}
    \caption{NLO prediction for the neutron relative-energy distribution with FSI (solid line) with EFT uncertainty estimated from the comparison of the NLO and LO (dashed line). All interaction parameters are as in Fig.~\ref{fig:NLOplot}.}
    \label{fig:NLOplot_errorband}
\end{figure}
In contrast to the ground-state distributions, the distribution with FSI shows a strong sensitivity to 
the $nn$ scattering length.  
One can devise a very conservative uncertainty band for the NLO result by assuming that the NLO uncertainty is equal to the difference between the LO and NLO results.\footnote{{\it A priori}, one would expect the uncertainty of the NLO result to be smaller than the difference of the LO and NLO results by a factor of the expansion parameter.}
The relative difference between the $a_{nn}=-18.7\,$fm LO and NLO curve is less than $5\%$ for energies $E_{nn} \leq 1\,$MeV and reaches about $10\%$ for $E_{nn} = 2\,$MeV. 
Below $1$\,MeV the behavior is non-monotonous because we normalize the distributions to one at $E_{nn}=0.8\,$MeV.
The corresponding uncertainty band is shown in Fig.~\ref{fig:NLOplot_errorband} for the case of $a_{nn}=-18.7$ fm.
Note that there are also uncertainties from the cutoff.
We use a cutoff of 700 MeV. Changing it to 1050 MeV, i.e., enlarging it by 50 \%, changes the \(nn\) distribution below energies of 2 MeV by up to about 1 \%, see Appendix \ref{ap:cutoff_convg} for the plot.
Also here the non-monotonous behavior due
to our normalization condition is observed.
Another calculation with a cutoff of 500 MeV shows a larger effect on the distribution and confirms the convergence pattern.
We conclude that the uncertainties from the cutoff are well under control and do not add much to the overall uncertainty.

\begin{figure}[htb]
    \centering
    \includegraphics[width=0.7\textwidth]{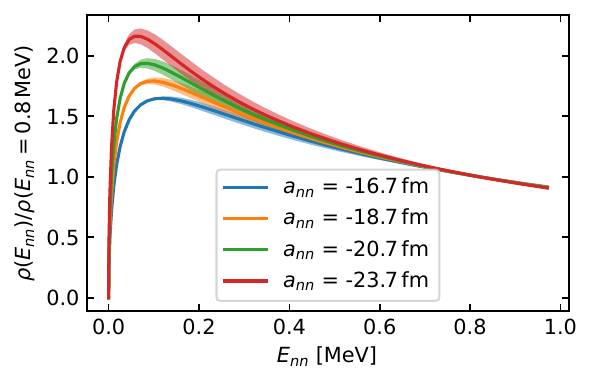}
    \caption{Neutron relative-energy distribution at NLO including FSI using the t-matrix method. Calculations use a cutoff of $\Lambda=700\,$MeV and the three-body force is tuned to reproduce $B_3=8.48\,$MeV.  All the curves have error bands based on the difference between the LO and NLO curves.}
    \label{fig:FSINLOerrorband}
\end{figure}
Our results for different scattering lengths $a_{nn}$ including the uncertainty bands are shown
in Fig.~\ref{fig:FSINLOerrorband} for the energy range 
up to $E_{nn}=1$ MeV relevant for the planned experiment.\footnote{Note that in the LO result for the uncertainty band, all effective ranges in the ground state and FSI contributions were consistently set to zero.}
The peak height as well as the peak position change with the scattering length. For example, the $a_{nn}=-16.7\,$fm curve has a peak at $E_{nn}=0.12\,$MeV with a height of $1.65$ whereas the $a_{nn}=-23.7\,$fm curve peaks at $E_{nn}=0.06\,$MeV with a height of $2.16$.
This means that the peak-to-tail ratio is very sensitive to variations of the scattering length.
Variations of the scattering length by $2$ fm around the value of $-18.7$ fm lead to changes in this ratio of about 10\%.
A plot of the distribution's sensitivity to the scattering length can be found in Appendix~\ref{ap:ann_sensitivity}.
We conclude that the neutron-neutron distribution 
in the hard knock-out reaction $t(p,2p)nn$ is a suitable 
observable to measure $a_{nn}$, similar to the neutron
distribution of the reaction $^6$He($p,p\alpha$)$nn$
investigated in \cite{Gobel:2021pvw}.
The conservative uncertainty bands of the curves for different scattering lengths are clearly separated in the peak region in all cases. 
In particular, the error bands for the two most relevant cases \(a_{nn} = -16.7\) fm and \(a_{nn} = -18.7\) fm corresponding  to the two classes of $a_{nn}$ values given in the literature (cf.~Refs.~\cite{Gardestig:2009ya,Gobel:2021pvw} for reviews) are well separated in the peak region.
This suggests that an NLO calculation for the neutron 
relative energy distribution of the reaction $t(p,2p)nn$ 
may be sufficient for the analysis of the planned experiment \cite{nn_scat_len_ribf_prop2018}.
However, an N$^2$LO calculation would be useful to further reduce the uncertainties and confirm
the expected convergence behavior of the effective field theory calculation.
Such calculation can also be carried out using the semi-perturbative
framework of Ref.~\cite{Bedaque:2002yg}.

Finally, note that we also used the technique of FSI enhancement factors to cross check our results.
This approximation gives consistent results to the exact
t-matrix method. Details of this comparison are given in Appendix \ref{ap:FSI_tm_ef}.

\section{Conclusion and outlook}
\label{sec:concl}

In this paper, we have calculated the two-neutron relative-energy distribution in the hard knock-out reaction $t(p,2p)nn$
at LO and NLO in pionless EFT.
Following Ref.~\cite{Bedaque:2002yg}, we included the effective ranges at NLO semi-perturbatively. Since we are primarily interested in the shape, we use a normalization scheme where the distribution is normalized to be one at some point (chosen as \(E_{nn}=0.8\) MeV).

We have started with the ground-state neutron relative-energy distribution at LO and NLO in pionless EFT.
The energy distribution raises sharply at zero energy, peaking around $E_{nn}=2.6\,$MeV, and then decreases slowly for higher energies. The relative deviation of the NLO result from the LO one is less than 1\,\% at relative energies of 1 MeV and 10\,\% at 10 MeV, smaller than the \textit{a priori} expectations from the power counting. 
The influence of the \(nn\) scattering length $a_{nn}$ on the ground-state distribution
is tiny for variations of $a_{nn}$ between -16.7 fm and -23.7 fm.
This lack of sensitivity can be understood from the 
closeness of $a_{nn}$ to the unitary limit as discussed in detail in Ref.~\cite{Gobel:2023rpj}. We note in passing, that a similar tiny influence of $a_{nn}$  on the ground-state distribution was also observed for the reaction $^6$He($p,p\alpha$)$nn$ \cite{Gobel:2021pvw}.

The next step was the calculation of the full distribution observed after the sudden knockout of the proton 
in the reaction $t(p,2p)nn$ as planned in the 
experiment at RIBF/RIKEN \cite{nn_scat_len_ribf_prop2018}.
Here it is crucial to include the FSI, i.e., the interactions of the fragments following the knockout. Because of the specific kinematics of a hard knock-out reaction, final-state interactions other than the \(nn\) FSI are suppressed.
We have included the \(nn\) FSI using the t-matrix approach
based on the triton ground-state wave functions at LO and NLO. 
We found that including the \(nn\) FSI changes the shape significantly.
In contrast to the ground-state distributions, there is a significant influence of the scattering length \(a_{nn}\) on the distribution.
Changing \(a_{nn}\) by 2 fm around the accepted value of 
$a_{nn}=-18.7$ fm causes variations of the distribution around its peak by about 10\,\%.

These observations are in line with those made for \(^6\)He \cite{Gobel:2021pvw}, where the FSI also caused much more a pronounced shape compared to the ground state and is almost exclusively the cause for sensitivity on \(a_{nn}\).
We conclude that the reaction $t(p,2p)nn$ is a suitable candidate for extracting the \(nn\) scattering length by fitting the theoretical shape to the measured shape.
As explained in Ref. \cite{nn_scat_len_ribf_prop2018}, having multiple nuclei suitable for this experiment provides an important cross check for the scattering-length measurement.
Similar to \(^6\)He in the reaction $^6$He$(p,p\alpha)nn$, the initial triton nucleus mainly serves as a neutron source for the experiment. Since the three-body force is readjusted to reproduce the physical binding energy of the triton when $a_{nn}$ is varied, it does not show any sensitivity to the exact value of $a_{nn}$. The window on the \(nn\) interaction is then provided by the final-state interaction.
Nevertheless, the measured shape is influenced by the ground-state structure and it is important to calculate its effect.

Based on the difference between the LO and NLO
results, we have generated conservative uncertainty bands for the NLO
calculation.
The EFT uncertainties at relative energies smaller 1 MeV reach up to about 5\,\%.
In the peak region they are of order 2\,\% or smaller and thus allow 
to separate the cases $a_{nn}=-16.7$ fm and $a_{nn}=-18.7$ fm, which differ by about 10\,\%. 
Therefore an analysis of the planned experiment based on an NLO calculation may be possible. 
Nevertheless, an extension of the pionless EFT calculation to 
N$^2$LO would be very valuable to further reduce the uncertainties and confirm the expected convergence behavior of the effective field theory calculation. At this order, no new two-body information is required, but another three-body force enters \cite{Bedaque:2002yg}.
The latter can, e.g., be determined from the spin-doublet neutron-deuteron scattering length.
Work in this direction is in progress.
Moreover, it would be interesting to explicitly calculate corrections from the kinematically suppressed final-state interactions (e.g. \(np\) FSI) in order to verify their kinematical suppression in the hard knock-out reaction.

Finally, the universality of \(nn\) distribution of various two-neutron halo nuclei has been studied in Ref.~\cite{Gobel:2023rpj}.
It would be interesting to investigate the distribution 
in the reaction $t(p,2p)nn$ calculated in the present paper from the perspective of universality.

\acknowledgments
We thank T. Aumann and D.R. Phillips for discussions.
T.K. and H.W.H. acknowledge support by Deutsche Forschungsgemeinschaft (DFG, German Research Foundation) - Project-ID 279384907 - SFB 1245.
H.W.H. has been supported by the German Federal Ministry of Education and Research (BMBF) (Grants No. 05P21RDFNB and 05P24RDB).

\appendix

\section{Further studies}
\subsection{Comparison of the t-matrix and enhancement-factor techniques}
\label{ap:FSI_tm_ef}

To cross-check the results for the $nn$ distributions with FSI, we compare the results based on the t-matrix approach, 
Eqs.~(\ref{eq:exactFSI}, \ref{eq:exactFSI-rho}), with those based on the enhancement-factor technique, Eqs.~(\ref{eq:enhancementfactor}, \ref{eq:enhancementfactor-rho}). We use a cutoff of $\Lambda=700\,$MeV, a $nn$ scattering length of $a_\mathrm{nn}=-18.7\,$fm, and the three-body force is tuned to reproduce the triton binding energy $B_3=8.48\,$MeV.
\begin{figure}[H]
    \centering
    \includegraphics[width=0.495\textwidth]{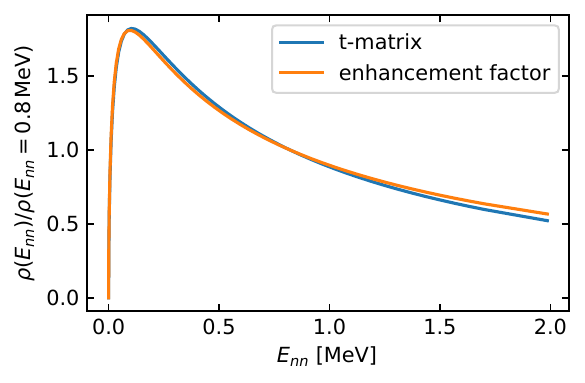}
    \includegraphics[width=0.495\textwidth]{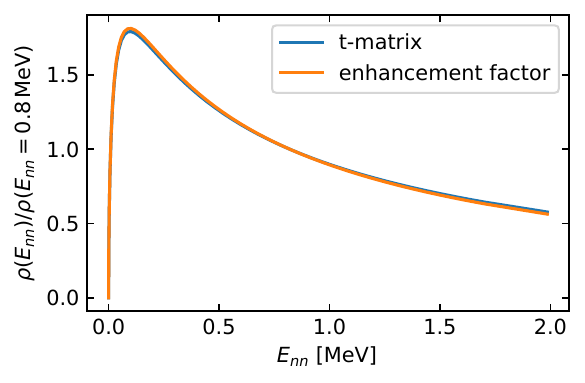}
    \caption{Two-neutron relative-energy distribution with FSI at NLO obtained for a cutoff of $\Lambda=700\,$MeV, a $nn$ scattering length of $a_\mathrm{nn}=-18.7\,$fm, and the three-body force tuned to reproduce $B_3=8.48\,$MeV. The FSI is included using the t-matrix method and the enhancement factor method. \textit{Left panel:} LO ground state with FSI ($r_{nn}=0$ for t-matrix method, $r_{nn}=2.84\,$fm for enhancement factor). \textit{Right panel:} NLO ground state and FSI calculations using the effective range $r_{nn}=2.84\,$fm.}
    \label{fig:FSI_comp}
\end{figure}

We find that the two FSI methods generally agree well. In particular, the agreement improves from LO (left) to NLO (right), where the two curves nearly lie on top of each other.
The slight disagreement between the two methods at LO is
probably due to the treatment of the 
neutron-neutron effective range in the enhancement factor. 
Since the enhancement factor, Eq.~\eqref{eq:enhancementfactor}, diverges as $r_{nn}\to 0$,
a purely LO calculation is not possible in this case.
Therefore, we have used the physical value $r_{nn}=2.84\,$fm for the enhancement factor.
At NLO, where the effective ranges are treated consistently,
the disagreement disappears and the two FSI methods agree 
very well.

\subsection{Convergence with cutoff}
\label{ap:cutoff_convg}

\begin{figure}[h]
    \centering
    \includegraphics[width=.7\textwidth]{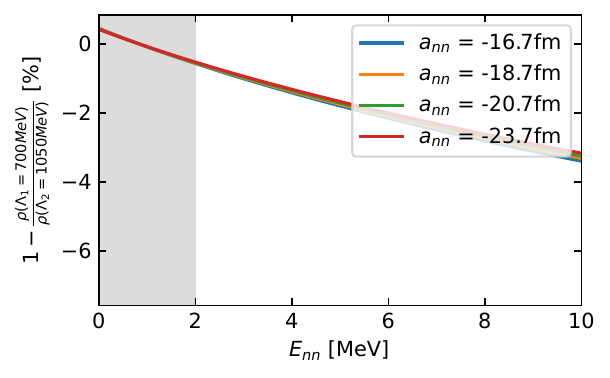}
    \caption{Relative deviations of NLO distribution obtained at \(\Lambda=700\) MeV from NLO distribution at \(\Lambda = 1050\) MeV in percent. The grey area marks the energy range on which we focus with respect to the experiment. The underlying distributions have FSI included via the t-matrix and have been normalized to be 1 at \(E_{nn} = 0.8\) MeV.}
    \label{fig:convergence_cutoff}
\end{figure}

We present details on the sensitivity of the results on the cutoff \(\Lambda\).
The relative deviation between the distribution at \(\Lambda = 700\) MeV and the one
at \(\Lambda = 1050\) MeV is shown in Fig.~\ref{fig:convergence_cutoff}.
Both distributions were obtained at NLO.

It can be seen that for relative energies below 2 MeV, the absolute values of the deviations
stay below 1 \%. The zero in this plot is due to normalizing the underlying distributions to 1 at \(E_{nn} = 0.8\) MeV.
As discussed in \cref{ssec:results_wfsi}, the uncertainty for the cutoff is already lower than the conservative estimate
for N2LO corrections based on the NLO-LO difference.
Moreover, the convergence pattern seen in our studies also shows that in future higher-order calculations the cutoff uncertainties can be still kept below the higher-order uncertainties by using a high-enough cutoff.

\subsection{Sensitivity on the scattering length}
\label{ap:ann_sensitivity}

\begin{figure}[h]
    \centering
    \includegraphics[width=0.7\textwidth]{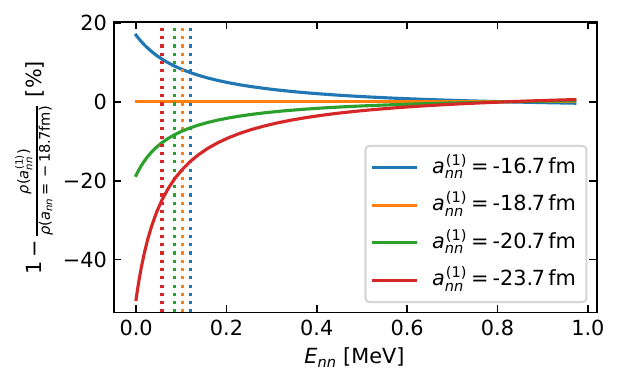}
    \caption{Sensitivity of the distribution on the scattering length based on NLO calculations with t-matrix FSI and the normalization to $1$ at $0.8\,$MeV. The dotted lines show the maxima of the final FSI curves calculated with the respective $nn$ scattering lengths.}
    \label{fig:sensitivity_ann}
\end{figure}

We evaluate the sensitivity on the $nn$ scattering length by calculating the relative difference between a distribution with some \(a_{nn}\) and the distribution calculated with $a_{nn}=-18.7\,$fm. This can be seen in Fig.~\ref{fig:sensitivity_ann}. The maxima of the respective curves are shown as vertical dotted lines. Note that the sensitivity naturally goes to 0 around 0.8 MeV due to the normalization scheme employed here, which allows us to focus on the distribution's shape. From the plot one can deduce the sensitivity of the peak-to-tail ratio of about 10 \% to variations of \(a_{nn}\) by 2 fm.

Comparing these curves to the relative differences of the respective enhancement factors shows good agreement. This is expected, given Appendix~\ref{ap:FSI_tm_ef} where we found good agreement between the results based on t-matrix FSI and those based on enhancement factors.

\bibliographystyle{apsrev4-1}
\bibliography{literature}

\end{document}